\documentclass[aps, prd, twocolumn, final, showpacs, nofootinbib, floatfix]{revtex4-1}

\usepackage{amsmath}
\usepackage{amssymb}
\usepackage{color}
\usepackage{verbatim}
\usepackage{graphicx}

\newcommand{\toponder}[1]{}
\newcommand{\mt}{\mathcal{T}}

\newcommand{\lag}{\mathcal{L}}
\newcommand{\tr}{\mathrm{tr}}
\newcommand{\hms}[3]{\ensuremath{{#1}^\text{h} {#2}^\text{m} {#3}^\text{s}}}
\newcommand{\dms}[3]{\ensuremath{{#1}^\circ {#2}' {#3}''}}

\newlength{\figwidth}
\begin{document}

\title{Direct-coupling lensing by antisymmetric tensor monopoles}

\author{Kamuela N.\ Lau} 
\email{Kamuela.N.Lau@williams.edu} 
\author{Michael D.\ Seifert} 
\email{Michael.Seifert@williams.edu}
\affiliation{Dept.\ of Physics, Williams College, 33 Lab Campus Dr.,
  Williamstown, MA 01267, USA}

\begin{abstract}

  We discuss the effects of a direct coupling between a rank-two
  antisymmetric tensor field and the Maxwell field.  The coupling we
  consider leads to vacuum birefringence, allowing us to place
  constraints on the magnitude of the tensor field and the strength of
  its coupling to the Maxwell field via cosmological birefringence
  measurements.  For light propagating in the presence of a
  topological defect solution, we find that light rays with different
  polarizations will follow different trajectories; the magnitude of
  this deflection is predicted to be extremely small (on the order of
  $10^{-10}$ arcseconds).  We discuss the plausibility of this
  phenomenon as a method for detection of these monopoles, along with
  the applicability of our methods to other possible couplings between
  the tensor field and the Maxwell field.

\end{abstract}

\pacs{}

\maketitle

\section{Introduction}

Since its discovery and development at the beginning of the 20th
century, Lorentz symmetry has been shown many times to be a very good
symmetry of nature.  Over the past decade and a half, however, there
has been significant interest in investigating the possible ways in
which Lorentz symmetry might be violated.  One of the more active
research programs being pursued to this end is the so-called
\emph{Standard Model Extension}, or SME \cite{Colladay1998}.  This
program ``extends'' the Standard Model by relaxing the requirement
that the field combinations appearing in the Standard-Model Lagrangian
be Lorentz scalars, and allowing combinations that are spacetime
tensors to appear in the Lagrangian.  Since the total action must
still be a scalar, this means that the coefficients of these new terms
will also be Lorentz tensors.  The presence of these terms in the
Lagrangian will affect the equations of motion of the fields (on the
classical level) and the field propagators and Feynman rules (on the
quantum level); in principle, then, these new tensor coefficients
could be measured experimentally.\footnote{Since these coefficients
  are tensors, their specification in terms of components is
  frame-dependent; the standard choice of frame, and the one we will
  use for this paper, is the ``Sun-centered frame'', in which the
  $z$-axis is parallel to the axis of Earth's rotation and the
  $x$-axis points towards the Vernal Equinox.  Spherical angles in
  these coordinates are the standard astronomical right ascension and
  declination.} While none of these postulated coefficients have thus
far been measured to be unambiguously non-zero, many bounds have been
placed on the values of these coefficients, some of which are
exceedingly stringent \cite{Kostelecky2008}.

The SME was initially developed from a particle-physics perspective,
and in particular, initially assumed that the ``arena'' in which
fields exist and interact was flat Minkowski spacetime.  In the
context of flat spacetime, it is legitimate to assume that these new
tensor coefficients are constants throughout space and time, and so
can be viewed as ``constants of nature''.  However, it was soon
realized that this simple picture runs into trouble when we try to
extend it to curved Riemannian spacetime.  The first obvious objection
in this case is that while many constant tensor fields exist on a flat
manifold (e.g., $\partial_a A^b = 0$ has many solutions), a constant
field (e.g., with $\nabla_a A^b = 0$) will not in general exist on a
curved manifold.  One might try to fix this problem by allowing the
tensor coefficients of the new terms to be fixed background tensor
fields in the case of curved spacetime.  However, this leads us to a
second problem with the original conception of the SME, less obvious
but more serious.  It was shown by Kosteleck\'y \cite{Kostelecky2004}
that the presence of a fixed background tensor field would necessarily
lead to violations of the Bianchi identities.  The only way to
circumvent this problem, while retaining a spacetime that is
well-described by Riemannian geometry, is to allow these ``background
tensor fields'' to be dynamic.  Schematically, the full action of the
theory will then be of the form
\begin{multline}
  S = \int d^4x \sqrt{-g} \left( - (\nabla \mt)(\nabla
    \mt) - V(\mt) \right. \\ \left. {} +
    \mathcal{L}_\text{SM}(\Psi)+ \mathcal{L}_\text{LV}(\Psi,
    \mt) \right),
\end{multline}
where $\mt$ is a ``Lorentz-violating'' tensor field and $\Psi$
represents the ``conventional'' matter fields of the Standard Model.
In this Lagrangian, $\mathcal{L}_\text{SM}$ is the Standard Model
Lagrangian (including gravity), while $\mathcal{L}_\text{LV}$
represents a small coupling between conventional matter and the
Lorentz-violating tensor field.  Since the Lorentz-violating tensor
fields will now be dynamic, they will satisfy their own
diffeomorphism-invariant equations of motion, and so the Bianchi
identities will automatically be satisfied so long as the equations of
motion for this field (and the conventional matter fields) hold.  By
appropriately choosing the form of the potential $V(\mt)$, we can
recover the ``original SME'' limit of flat spacetime and a constant
background tensor field coupled to conventional matter.  In such a
picture, Lorentz symmetry is broken spontaneously rather than
explicitly. 

Since the Lorentz-violating tensor field $\mt$ must be dynamical, it
is natural to ask how this field will behave.  The dynamics of such
tensor fields in flat spacetime have been widely studied over the past
few years, particularly in the case where $\mt$ is a vector field
$A^a$ \cite{Bluhm2005, Bluhm2008, Seifert2009, Seifert2010b}, a
symmetric rank-two tensor field $C^{ab}$ \cite{Kostelecky2009}, or an
antisymmetric rank-two tensor field $B_{ab}$ \cite{Altschul2010}.
Most of this prior work has centered on the situation where the tensor
field is constant throughout space, taking on some value $\bar{\mt}$
that minimizes its potential energy.  However, this field value will
not generally speaking be unique.  A general potential
$V(\mathcal{T})$ which breaks Lorentz symmetry will possess many
possible values of $\mathcal{T}$ which minimize the potential energy;
these field values form the so-called \emph{vacuum manifold} in field
space.  This raises the possibility of topological defect solutions in
the cases where the vacuum manifold is topologically non-trivial, and
it has been shown that in the case of the antisymmetric rank-two
tensor field, there exist topological monopole solutions
\cite{Seifert2010, Seifert2010a}.  These solutions are static, stable,
and spherically symmetric, but the field is not in the vacuum manifold
except asymptotically, and approaches different points in the vacuum
manifold as we go to spatial infinity in different directions.
Moreover, if an antisymmetric tensor field capable of supporting these
defects exists, one would expect a certain number of these monopoles
to arise in the early Universe via the Kibble mechanism
\cite{Kibble1976}, and these monopoles might persist as relics today.
In other words, not only are non-constant tensor fields required in
curved spacetimes, but they arise naturally in the context of flat
spacetime as well.

The possibility of a stable non-constant background for a
Lorentz-violating field leads us to ask what the effects of such a
field on matter would be.  To the best of our knowledge, all prior
work dealing with Lorentz violation (particularly within the SME
program \cite{Bluhm2005, Bluhm2008, Seifert2009,
  Seifert2010b,Kostelecky2009,Altschul2010}) has only dealt with
position-independent effects; however, the existence of monopole
solutions provides us with the motivation that position-dependent
effects might be physically relevant.  In this paper, we examine the
effects of an antisymmetric tensor monopole as a background field
coupled to the electromagnetic field.  In Section \ref{LVEsec}, we
present our model and review the formalism used to examine
Lorentz-violating effects in electrodynamics.  In Section
\ref{paramboundsec}, we use polarimetry data to gain an idea of the
size of the effects which will arise and bound the parameters of our
model.  The main results of this work are then derived in Section
\ref{lensingsec}, in which it is shown that an antisymmetric tensor
monopole coupled directly to light would cause a lensing effect on
light rays, akin to a gravitational lensing effect but in flat
spacetime.

We will use units throughout where $\hbar = c = 1$.  The metric
signature will be $(-, +, +, +)$.

\section{Lorentz-violating electrodynamics}
\label{LVEsec}

The main objects of interest in this paper will be a conventional
Maxwell field $A_a$ and an antisymmetric rank-two tensor field
$B_{ab}$. The action for this model is
\begin{equation}
  \lag = - \frac{1}{4} F_{ab} F^{ab} - \frac{1}{6} F_{abc} F_{abc} -
  V(B_{ab}) - (k_F)^{abcd} F_{ab} F_{cd},
  \label{lagdef}
\end{equation}
where $F_{ab} = 2 \partial_{[a} A_{b]}$ is the field strength for the
Maxwell field;  $F_{abc} = 3 \partial_{[a} B_{bc]}$ is the field
strength for the ``Lorentz-violating'' field $B_{ab}$;  $V(B_{ab})$ is
a potential energy for $B_{ab}$, given by
\begin{equation}
  V(B_{ab}) = \frac{\lambda}{4} \left( B_{ab} B^{ab} - b^2 \right)^2;
  \label{potdef}
\end{equation}
and $(k_F)^{abcd}$ is a tensor that couples the Maxwell field to the
Lorentz-violating field, given by\footnote{This is, of course, not the
  only possible coupling between $B_{ab}$ and $A_a$; see Section
  \ref{discsec} for further discussion.}
\begin{equation}
  (k_F)^{abcd} = \xi \left( B^{ab} B^{cd} -  B^{[ab} B^{cd]} -
    \frac{1}{6} \eta^{a[c} \eta^{d] b} B_{ef} B^{ef} \right).
  \label{kdef}
\end{equation}
The tensor $(k_F)^{abcd}$ can be seen to be dimensionless; this
implies that the parameter $\xi$ has mass dimension $-2$. The
parameter $\xi$ is assumed to be a ``small'' parameter that determines
the strength of the coupling; we will assume it to be non-zero, but in
principle it could be either positive or negative.  The first
``subtraction'' term in \eqref{kdef} is chosen to eliminate a term
proportional to $\epsilon_{abcd} F^{ab} F^{cd}$, which is a total
derivative and so would not contribute to the equations of motion.
The second subtraction is chosen to eliminate a term proportional to
$F_{ab} F^{ab}$, which would simply rescale the factor of
$\frac{1}{4}$ in front of the ``Lorentz-invariant'' kinetic term for
the Maxwell field.  An arbitrary tensor with these symmetries would
have 19 independent components \cite{Kostelecky2002}; however, since
our $(k_F)^{abcd}$ is constructed out of the antisymmetric tensor
$B_{ab}$, it only has six independent degrees of freedom.

The effects of an arbitrary tensor $(k_F)^{abcd}$ on electromagnetic
fields have been formalized by Kosteleck\'y and Mewes
\cite{Kostelecky2002}.  The behaviour of the fields in the presence of
such a coupling is equivalent to their behaviour in a linear medium:
\begin{equation}
  \begin{aligned}
    \vec{\nabla} \cdot \vec{D} &= 0, & \vec{\nabla} \times \vec{H} -
    \frac{\partial \vec{D}}{\partial t} &= 0, \\
    \vec{\nabla}\cdot \vec{B} &= 0, & \vec{\nabla} \times \vec{E} +
    \frac{\partial \vec{B}}{\partial t} &= 0.
  \end{aligned}
  \label{modmax}
\end{equation}
The constitutive relations of this ``medium'' are given by
\begin{equation}
  \begin{bmatrix} \Vec{D} \\ \Vec{H} \end{bmatrix} = \begin{bmatrix} \mathbf{1} +
    \kappa_{DE} & \kappa_{DB} \\ \kappa_{HE} & \mathbf{1} + \kappa_{HB}
  \end{bmatrix} \begin{bmatrix} \Vec{E} \\ \Vec{B} \end{bmatrix},
  \label{constitutive}
\end{equation}
where $\mathbf{1}$ is the $3 \times 3$ identity matrix and the
$\kappa$'s are $3 \times 3$ matrices given by
\begin{subequations}
  \begin{align}
    (\kappa_{DE})^{ij} &= -2 (k_F)^{0i0j}, \\
    (\kappa_{DB})^{ij} = - (\kappa_{HE})^{ji} &= (k_F)^{0ikl}
    \epsilon^{jkl}, \text{ and} \\
    (\kappa_{HB})^{ij} &= \frac{1}{2} \epsilon^{ikl} \epsilon^{jmn}
    (k_F)^{klmn}.
    \end{align}
\end{subequations}
We define the spatial vectors $\Vec{Q}$ and $\Vec{R}$ to be the
``electric'' and ``magnetic'' parts of $B^{ab}$, respectively; in
other words, $Q^i = B^{0i}$ and $R^i = \frac{1}{2} \epsilon^{ijk}
B^{jk}$.  In terms of these vectors, we find that
\begin{subequations}
  \label{kappamats}
  \begin{align}
    (\kappa_{DE})^{ij} &= \xi \left[ - 2 Q^i Q^j + \frac{1}{3}
      (\Vec{Q}^2 - \Vec{R}^2) \delta^{ij} \right], \\
    (\kappa_{DB})^{ij} &= \xi \left[ -2 Q^i R^j + \frac{2}{3} (\Vec{Q}
      \cdot \Vec{R}) \delta^{ij} \right], \text{ and} \\
    (\kappa_{HB})^{ij} &= \xi \left[ 2 R^i R^j + \frac{1}{3} (
      \Vec{Q}^2 - \Vec{R}^2 ) \delta^{ij} \right].
  \end{align}
\end{subequations}
We can see immediately from \eqref{kappamats} that the ``linear
medium'' for our electromagnetic fields will be, in general,
anisotropic; in fact, the only way for all of these matrices to be
isotropic (i.e., proportional to $\delta^{ij}$) is for all of them to
vanish, with $\Vec{Q} = \Vec{R} = 0$.

These $\kappa$ matrices can also be reparametrized in terms of certain
parity-even and parity-odd combinations, denoted $\tilde{\kappa}_{e
  \pm}$, $\tilde{\kappa}_{o\pm}$, and $\tilde{\kappa}_\tr$
\cite{Kostelecky2002}; experimental bounds on the components of the
tensor $(k_F)^{abcd}$ are usually quoted in terms of these matrices
\cite{Kostelecky2008}.  In our case, we have 
\begin{subequations}
  \label{tkmats}
  \begin{align}
    (\tilde{\kappa}_{e\pm})^{ij} &= \xi \left[ - Q^i Q^j \pm R^i R^j +
      \frac{1}{3} (\Vec{Q}^2 \mp \Vec{R}^2) \delta^{ij} \right], \\
    (\tilde{\kappa}_{o+})^{ij} &= 2 \xi Q^{[i} R^{j]}, \\
    (\tilde{\kappa}_{o-})^{ij} &= 2 \xi \left[ - Q^{(i} R^{j)} +
      \frac{1}{3} (\Vec{Q} \cdot \Vec{R}) \delta^{ij} \right], \text{ and} \\
    \tilde{\kappa}_\tr &= - \frac{\xi}{3} (\Vec{Q}^2 + \Vec{R}^2).
  \end{align}
\end{subequations} 
Since in all of the above expressions a change in the coupling
strength $\xi$ is indistinguishable from a simultaneous rescaling of
$\Vec{Q}$ and $\Vec{R}$, we will define two new vectors $\bar{Q} =
\sqrt{|\xi|} \Vec{Q}$ and $\bar{R} = \sqrt{|\xi|} \Vec{R}$. In terms
of these rescaled vectors, we have 
\begin{subequations}
  \label{tkmatsrescaled}
  \begin{align}
    (\tilde{\kappa}_{e\pm})^{ij} &= \bar{\xi} \left[ - \Bar{Q}^i
      \Bar{Q}^j \pm \Bar{R}^i \Bar{R}^j +
      \frac{1}{3} (\bar{Q}^2 \mp \bar{R}^2) \delta^{ij} \right], \\
    (\tilde{\kappa}_{o+})^{ij} &= 2 \bar{\xi} \Bar{Q}^{[i} \Bar{R}^{j]}, \\
    (\tilde{\kappa}_{o-})^{ij} &= 2 \bar{\xi} \left[ - \Bar{Q}^{(i} \Bar{R}^{j)} +
      \frac{1}{3} (\bar{Q} \cdot \bar{R}) \delta^{ij} \right], \text{ and} \\
    \tilde{\kappa}_\tr &= - \frac{\bar{\xi}}{3} (\bar{Q}^2 + \bar{R}^2),
  \end{align}
\end{subequations}  
where $\bar{\xi} = \xi/|\xi| = \pm 1$. 

\section{Constant-field parameter bounds}
\label{paramboundsec}

In almost all of the literature on Lorentz symmetry violation to date,
it is assumed that the Lorentz-violating fields are constant in space,
with a possibility of small linearized perturbations about a constant
background.  The monopole backgrounds that we will be examining are
fundamentally non-linear, and so some of the effects we will be
examining do not have an analog in the current literature.  However,
we can still examine the behaviour of our model in the case of a
constant background field in order to get an idea of the size of the
effects we are looking for.  
   
It is known that if either $\tilde{\kappa}_{e+}$ or
$\tilde{\kappa}_{o-}$ is non-zero, electromagnetic waves will
experience vacuum birefringence \cite{Kostelecky2002,
  Kostelecky2001}. We can see from the form of \eqref{tkmatsrescaled}
that a coupling of the form \eqref{kdef} will always produce
birefringence unless both $\bar{Q}$ and $\bar{R}$ are
zero.  Non-observance of vacuum birefringence will thereby allow us to
place stringent bounds on the components $\Bar{Q}^i$ and $\Bar{R}^i$,
which will in turn allow us to estimate the order of magnitude of the
parameter combination $\sqrt{|\xi|} b$. As we shall see, it is this
last combination of parameters that will determine the magnitude of
the light-bending effects that are the primary concern of this paper.

To do this, we apply the results of the analysis of Kosteleck\'y and
Mewes in \cite{Kostelecky2001}. Their analysis was able to place
bounds on the magnitude of parameter $\sigma$ (equal to twice the
difference in the phase velocity between the two polarizations) for a
list of sixteen optical and infrared sources, shown in Table
\ref{sourcetable}.
\begin{table}[tb]
  \begin{tabular}{lcrc}
    \multicolumn{1}{c}{Source}& $\alpha$ &
    \multicolumn{1}{c}{$\delta$}  & $\log_{10} |\sigma|$ \\ \hline
    IC 5063 & \hms{20}{52}{02} & \dms{-57}{04}{08}& -30.8  \\
    3A 0557-383 & \hms{05}{58}{02} & \dms{-38}{20}{04} & -31.2 \\
    IRAS 18325-5925 & \hms{18}{36}{58} & \dms{-59}{24}{08} & -31.0
    \\
    IRAS 19850-1818 & \hms{20}{00}{52} & \dms{-18}{10}{27} & -31.0
    \\
    3C 324 & \hms{15}{49}{49} & \dms{21}{25}{39} & -32.2 \\
    3C 256 & \hms{11}{20}{43} & \dms{23}{27}{55} & -32.2 \\
    3C 356 & \hms{17}{24}{19} & \dms{50}{57}{40} & -32.2 \\
    FIRST J084044.5+363328 & \hms{08}{40}{45} & \dms{36}{33}{28} &
    -32.2 \\
    FIRST J155633.8+351758 & \hms{15}{56}{34} & \dms{35}{17}{57} &
    -32.2 \\
    3CR 68.1 & \hms{02}{32}{29} & \dms{34}{23}{46} & -32.2 \\
    QSO J2359-1241 & \hms{23}{59}{54} & \dms{-12}{41}{48} & -31.1 \\
    3C 234 & \hms{10}{01}{50} & \dms{28}{47}{09} & -31.7 \\
    4C 40.36 & \hms{18}{10}{56} & \dms{40}{45}{24} & -32.2 \\
    4C 48.48 & \hms{19}{33}{05} & \dms{48}{11}{42} & -32.2 \\
    IAU 0211-122 & \hms{02}{14}{17} & \dms{-11}{58}{45} & -32.2 \\
    IAU 0828+193 & \hms{08}{30}{53} & \dms{19}{13}{16} & -32.2
  \end{tabular}
  \caption{\label{sourcetable} Sources used in the bounds placed in
    Section \ref{paramboundsec}.  Bounds on $|\sigma|$ are those from
    \cite{Kostelecky2001};  right ascension $\alpha$ and declination
    $\delta$ were found in the Centre de Donne\'{e}s astronomiques de
    Strasbourg online catalog (\texttt{http://cdsweb.u-strasbg.fr/}). }
\end{table}
This parameter $\sigma_A$ for a given source $A$ can in turn be
expressed in terms of the components of the tensor $(k_F)^{abcd}$ and
the right ascension and declination of the source $\{ \alpha_A,
\delta_A \}$.  Putting all of these together, then, we see that each
independent bound on $\sigma_A$ will constrain some polynomial
function of the $\Bar{Q}^i$ and $\Bar{R}^i$, and that this function
will depend on the right ascension and declination of the source in
question.  The actual form of these functions is quite complicated,
and the reader is referred to the Appendix for details on how they are
constructed.

We treat each of these sixteen two-sided bounds in
\cite{Kostelecky2001} as a strict exclusion; this leaves a small
region of parameter space near the origin that is allowed under the
simultaneous imposition of all of the bounds.  The maximum magnitudes
of the components $\bar{Q}^x$, $\bar{Q}^y$, and $\bar{Q}^z$ in this
allowed region of parameter space are $1.56 \times 10^{-16}$, $1.54
\times 10^{-16}$, and $1.43 \times 10^{-16}$ respectively.  The bounds
on the magnitudes of $\bar{R}^x$, $\bar{R}^y$, and $\bar{R}^z$ are
identical; this is to be expected, since all of the coefficients that
control birefringent effects (i.e., the components of
$\tilde{\kappa}_{e+}$ and $\tilde{\kappa}_{o-}$) are antisymmetric
under the exchange $\Vec{Q} \leftrightarrow \Vec{R}$ and all of our
observational bounds are taken to be symmetric about zero.  Finally,
the value of $\xi B_{ab} B^{ab} = 2 \xi ( - \Vec{Q}^2 + \Vec{R}^2)$ in
the allowed region of parameter space is bounded by
\begin{equation}
  |\xi B_{ab} B^{ab}| < 4.59 \times 10^{-32}.
\end{equation}
This bound is symmetric about zero for the same reasons that the
individual bounds on the components of $\bar{Q}$ and
$\bar{R}$ are the same.

During the course of the preparation of this work, another work
\cite{Kostelecky2013} was published, based on six polarization
measurements of gamma-ray bursts.  The bounds on certain components of
$(k_F)^{abcd}$ derived in this latter work were much more stringent
than those from the work \cite{Kostelecky2001}, by a factor of
approximately $10^6$. However, the bounds on the coefficients of
$(k_F)^{abcd}$ in both \cite{Kostelecky2001} and \cite{Kostelecky2013}
are derived under the assumption that the light is not
``accidentally'' emitted in a normal mode, in which case the
birefringent effects would be unobservable.  Since our parameter space
is effectively six-dimensional, and only six sources were used in the
derivation of these newer bounds to begin with, the loss of any one of
the newer bounds in \cite{Kostelecky2013} would still leave a
parameter of our model effectively unconstrained.  The larger number
of data points used in the earlier work \cite{Kostelecky2001} leads to
more robust constraints, since deletion of one or two data points
would not substantially affect the bounds.  In principle, a larger set
of measurements of the type and precision found in
\cite{Kostelecky2013} could be subjected to the same analysis, and one
would expect that such an analysis would lead to bounds on $\xi B_{ab}
B^{ab}$ that were proportionally more stringent.

\section{Monopole lensing}
\label{lensingsec}
\subsection{Ordinary \& extraordinary waves}

In the previous section, we worked under the assumption that the
background field $B_{ab}$ was constant in space and time.  However,
recent work on monopole solutions \cite{Seifert2010, Seifert2010a,
  Li2012} has shown that there exist static solutions in which the
field $B_{ab}$ varies spatially, both in magnitude and direction.  In
the context of Lorentz-violating electrodynamics, this means that the
$\kappa$ matrices appearing in the constitutive relations
\eqref{constitutive} vary from point to point.  In other words, in the
presence of a monopole solution, electromagnetic fields act as though
they were in an inhomogeneous, anisotropic, birefringent medium.  In
particular, the inhomogeneity of the medium implies that light rays
travelling near the monopole will be deflected from straight-line
paths.  

To quantify this effect, we apply a geometric-optics approximation to
the modified Maxwell equations \eqref{modmax}.  Our derivation will
roughly follow the technique of Sluijter \emph{et
  al.}~\cite{Sluijter2008} where applicable.\footnote{Note that in
  Sluijter \emph{et al.}~\cite{Sluijter2008}, the medium of
  propagation is assumed to be electrically anisotropic but
  magnetically isotropic.  In the present work, however, our
  ``medium'' will turn out to be electrically isotropic but
  magnetically anisotropic. To apply the results of Sluijter \emph{et
    al.}~to our situation, then, we must effectively switch the roles
  of $\vec{E}$ and $\vec{H}$. \label{ehfootnote}} We choose an ansatz
for the electromagnetic fields of the form
\begin{subequations}
  \label{ebansatz}
  \begin{align}
    \Vec{E}(t,\Vec{x}) &= \Vec{e}(\Vec{x}) e^{ik(S(\Vec{x}) - t)}, \\
    \Vec{B}(t,\Vec{x}) &= \Vec{b}(\Vec{x}) e^{ik(S(\Vec{x}) - t)}.
  \end{align}
\end{subequations}
Plugging these into the modified Maxwell equations, we obtain the
equations
\begin{subequations}
  \label{modmaxansatz}
  \begin{align}
    \Vec{\nabla} S \times \Vec{h} + \Vec{d} &= - \frac{1}{ik}
    \Vec{\nabla} \times \Vec{h} \label{modampansatz} \\
    \Vec{d} \cdot \Vec{\nabla} S &= - \frac{1}{ik} \Vec{\nabla} \cdot
    \Vec{d} \\
    \Vec{b} \cdot \Vec{\nabla} S &= - \frac{1}{ik} \Vec{\nabla} \cdot
    \Vec{b} \\
    \Vec{\nabla} S \times \Vec{e} - \Vec{b} &= -\frac{1}{ik}
    \Vec{\nabla} \times \Vec{e}, \label{modfaradansatz}
  \end{align}
\end{subequations}
where we have defined
\begin{align}
  \Vec{d} &= (\mathbf{1} + \kappa_{DE}) \Vec{e} + \kappa_{DB}
  \Vec{b}, \label{littleddef} \\
  \Vec{h} &= \kappa_{HE} \Vec{e} + (\mathbf{1} + \kappa_{HB})
  \Vec{b}. \label{littlehdef}
\end{align}
(Recall that the $\kappa$'s in these equations are $3 \times 3$
matrices acting on the vectors $\Vec{e}$ and $\Vec{b}$.)  We then
apply the standard geometric-optics approximation, and restrict our
attention to solutions for which the length scale variation of the
vectors $\Vec{e}$, $\Vec{b}$, $\Vec{d}$, and $\Vec{h}$ are much less
than the length scale defined by $k$.  In other words,
\begin{equation}
  \left\{ \frac{ |\Vec{\nabla} \Vec{e}|}{|\Vec{e}|}, \frac{
      |\Vec{\nabla} \Vec{b}|}{|\Vec{b}|}, \frac{ |\Vec{\nabla}
      \Vec{d}|}{|\Vec{d}|}, \frac{ |\Vec{\nabla} \Vec{h}|}{|\Vec{h}|}
  \right\} \ll k
\end{equation}
in some appropriate sense, which allows us to neglect the right-hand
sides of the four equations \eqref{modmaxansatz}.  Note that if the
length-scale of variation of the $\kappa$ matrices is ``small'' in
this sense, then the slow variation of $\Vec{d}$ and $\Vec{h}$ follows
from the slow variation of $\Vec{e}$ and $\Vec{b}$.  

So far in this section, we have not assumed that the $\kappa$ matrices
have any particular form; our equations are valid for any (static)
background field $B_{ab}$, assuming it is static and slowly varying.
For the case of the monopole solution originally found in
\cite{Seifert2010}, the field configuration is of the form
\begin{equation}
  B_{\theta \phi} = g(r) r^2 \sin \theta, \label{bansatz}
\end{equation}
with all other components vanishing.  The function $g(r)$ is the solution
to the differential equation\footnote{This equation is just the
  equation of motion for $B_{ab}$ derived from the Lagrangian
  \eqref{lagdef}, under the imposition of the ansatz \eqref{bansatz}.}
\begin{equation}
  \frac{\partial}{\partial r} \left( \frac{\partial g}{\partial r} +
    \frac{2}{r} g \right) - 2 \lambda (2 g^2 - b^2) g = 0
\end{equation}
subject to the boundary conditions $g(0) = 0$ and $g(\infty) =
b/\sqrt{2}$.  While a closed-form solution for $g(r)$ is not known,
its asymptotic behaviour as $r \to \infty$ is
\begin{equation}
  g(r) = \frac{b}{\sqrt{2}} \left( 1 - \frac{1}{4 \lambda b^2 r^2} -
    \frac{3}{8 \lambda^2 b^4 r^4} + \dotsb \right).
  \label{gseries}
\end{equation}

In terms of the ``electric'' and ``magnetic'' vectors $\Vec{Q}$ and
$\Vec{R}$, it can easily be shown that for this solution we have
$\Vec{Q} = 0$ and 
\begin{equation}
  \Vec{R} = g(r) \hat{r}.
\end{equation}
From Equation \eqref{kappamats}, we see that the ``off-diagonal''
matrices $\kappa_{DB}$ and $\kappa_{HE}$ vanish, and that
$\kappa_{DE}$ becomes isotropic:
\begin{align}
  (\kappa_{DE})^{ij} &= - \frac{1}{3} \xi g^2(r) \delta^{ij} \\
  (\kappa_{HB})^{ij} &= \xi g^2(r) \left( 2 \hat{r}^i \hat{r}^j -
    \frac{1}{3} \delta^{ij} \right).
\end{align}
This allows us to define effective permittivity and permeability
tensors $\epsilon^{ij}$ and $\mu^{ij}$ such that $d^i = \epsilon^{ij}
e^j$ and $b^i = \mu^{ij} h^j$.  The permittivity tensor will be 
\begin{equation}
  \epsilon^{ij} = \delta^{ij} + (\kappa_{DE})^{ij} = \left( 1 -
    \frac{1}{3} \xi g^2 \right) \delta^{ij},
\end{equation}
while the inverse of the permeability tensor will be
\begin{equation}
  (\mu^{-1})^{ij} = \delta^{ij} + (\kappa_{HB})^{ij} = \left( 1 -
    \frac{1}{3} \xi g^2 \right) \delta^{ij} + 
  2 \xi g^2 \Hat{r}^i \Hat{r}^j.
\end{equation}
This last matrix can be inverted to yield
\begin{equation}
  \mu^{ij} = \frac{1}{1 - \frac{1}{3} \xi g^2} \left[ \delta^{ij} -
    \frac{2 \xi g^2}{1 + \frac{5}{3} \xi g^2} \hat{r}^i \hat{r}^j
  \right].
\end{equation}
In the language of birefringent optics, then, the ``medium'' in which
our waves are propagating will be electrically isotropic, magnetically
anistropic, and unaxial; the ``optical axis'' of the medium will be
the radial direction $\hat{r}$.

Combining all of the above, then, we have from \eqref{modampansatz}
and \eqref{littleddef}
\begin{equation}
  \vec{e} = - \frac{1}{\epsilon} (\Vec{\nabla} S) \times \vec{h},
  \label{ephrel}
\end{equation}
where $\epsilon = 1 - \frac{1}{3} \xi g^2$ is the (position-dependent)
permittivity.  Plugging this in to \eqref{modfaradansatz}, and
defining the wave-normal vector $\vec{p} = \vec{\nabla} S$, we obtain
\begin{equation}
  \left[ p^i p^j - (\Vec{p}\cdot\vec{p}) \delta^{ij} +
    \epsilon \mu^{ij} \right] h^j = 0.
  \label{propmatrix}
\end{equation}
For a non-trivial wave amplitude $\vec{h}$ to exist, it must be the
case that
\begin{equation}
  \det  \left[ p^i p^j - p^2 \delta^{ij} +
    \epsilon \mu^{ij} \right] = 0;
  \label{propdeterminant}
\end{equation}
after some algebra, this condition boils down to
\begin{equation}
  (p^2 - 1)\left( p^2 - \frac{\zeta}{1 + \zeta} (\vec{p}
    \cdot \hat{r})^2 - \frac{1}{1 + \zeta} \right) = 0, 
  \label{hamiltonian}
\end{equation}
where
\begin{equation}
  \zeta = \frac{2 \xi g^2}{1 - \frac{1}{3} \xi g^2}.
  \label{bdef}
\end{equation}

For a given direction of $\hat{r}$ and value of $\zeta$, the equation
\eqref{hamiltonian} defines a surface in $\vec{p}$-space called the
\emph{optical indicatrix}.  This surface consists of two ellipsoids
which are tangent at the points $\vec{p} = \pm \hat{r}$, and which do
not otherwise intersect.  These ellipsoids correspond to ordinary and
extraordinary waves, respectively; the ordinary waves satisfy
\begin{equation}
  \mathcal{H}_o = \frac{1}{2}\left( p_o^2 - 1 \right) = 0,
  \label{ordhamdef}
\end{equation}
while the extraordinary waves satisfy
\begin{equation}
  \mathcal{H}_e = \frac{1}{2}\left( (1 + \zeta) p_e^2 -
    \zeta (\vec{p}_e \cdot \hat{r})^2 - 1 \right) = 0.
  \label{extraordhamdef}
\end{equation}
(The reason for the choice of the overall factors in these equations
will become clear in Section \ref{raytrace}.) 

For the ordinary waves, requiring that $\vec{p}_o = \hat{p}_o$ is a
unit vector causes Equation \eqref{propmatrix} to reduce to
\begin{equation}
  \left[ \hat{p}_o^i \hat{p}_o^j - \frac{\zeta}{1 + \zeta} \hat{r}^i \hat{r}^j \right] 
  h^j_o = 0,
\end{equation}
It is not hard to see that for this to be true, the vector $\vec{h}_o$
must be at right angles to both $\hat{p}_o$ and $\hat{r}$.\footnote{In
  the case where $\hat{p}_o \parallel \hat{r}$, the wave direction is
  parallel to the optical axis, and so the ordinary and extraordinary
  waves travel at the same speed.}  In other words, we can define the
magnetic field direction $\hat{h}_o$ as
\begin{equation}
  \hat{h}_o = \frac{\hat{p}_o \times \hat{r}}{|\hat{p}_o \times
    \hat{r}|}.
  \label{ordmagpol}
\end{equation}
Equation \eqref{modampansatz} then tells us that the (electric)
polarization of the ordinary wave $\hat{e}_o$ is given by
\begin{equation}
  \hat{e}_o = \frac{ \hat{r} - \hat{p}_o (\hat{p}_o \cdot \hat{r}) }{|
    \hat{r} - \hat{p}_o (\hat{p}_o \cdot \hat{r})| }.
  \label{ordelecpol}
\end{equation}
In other words, the electric polarization for an ordinary wave points
along the projection of $\hat{r}$ in the plane orthogonal to
$\hat{p}_o$.  These waves propagate at constant speed through space,
since $p^2_o$ is constant.

The polarization of the extraordinary waves is somewhat less obvious.
We shall simply cite the work of \cite{Sluijter2008}; with the
substitution $\Vec{E} \leftrightarrow \Vec{H}$ (as noted in Footnote
\ref{ehfootnote}), the magnetic field direction $\hat{h}_e$ for an
extraordinary wave with wave front vector $\vec{p}_e$ can be seen to
point in the direction
\begin{equation}
  \hat{h}_e = \frac{(\vec{p}_e \times \hat{r}) \times [(1 + \zeta) \vec{p}_e -
    \zeta (\vec{p}_e \cdot \hat{r})  \hat{r} ]}{|(\vec{p}_e
    \times \hat{r}) \times  [(1 + \zeta) \vec{p}_e -
    \zeta (\vec{p}_e \cdot \hat{r})  \hat{r} ]|}, \label{extordh}
\end{equation}
while the electric field vector $\hat{e}_e$ will point in the
direction
\begin{equation}
  \hat{e}_e = - \frac{\vec{p}_e \times \hat{r}}{|\vec{p}_e \times
    \hat{r}|}.  \label{extorde}
\end{equation}
Note that the magnetic polarization direction $\hat{h}_e$ is not, in
general, perpendicular to the extraordinary wave-front vector
$\vec{p}_e$.

\subsection{Ray-tracing}
\label{raytrace}

To trace the paths of rays in the presence of a monopole background,
we follow the method of Sluijter \emph{et al.}~\cite{Sluijter2008} and
interpret the factors $\mathcal{H}_o$ and $\mathcal{H}_e$ as
point-particle Hamiltonians, whose trajectories evolve with respect to
some parameter $\tau$.  In other words, we expect that for both
ordinary rays and extraordinary rays, their position $\vec{x}(\tau)$
and momentum $\vec{p}(\tau)$ will satisfy
\begin{align}
  \frac{d \vec{x}}{d\tau} &=  \nabla_{\vec{p}} \mathcal{H} &
  \text{and}&& \frac{d \vec{p}}{d\tau} &= - \nabla_{\vec{x}}
  \mathcal{H}.
\end{align}
Under this interpretation, it is evident that the ordinary rays will
travel in straight lines:  the ordinary Hamiltonian $\mathcal{H}_o$ is
simply that of a free particle, and we will have
\begin{align}
  \frac{d \vec{x}}{dt} &= \vec{p} &\text{and}&& \frac{d \vec{p}}{dt}
  &= 0.
\end{align}
(The insertion of the extra factors of $\frac{1}{2}$ in Equations
\eqref{ordhamdef} and \eqref{extraordhamdef} was done to make these
equations look ``nice''.)  Thus, the velocity $\vec{v}_o$ of an
ordinary ray is constant with respect to time.  Its electric and
magnetic polarization directions will also be constant along its path:
since the velocity is parallel to the wave-normal direction
$\hat{p}_o$, it can be seen from \eqref{ordmagpol} and
\eqref{ordelecpol} that $\vec{v}_o$, $\hat{h}_o$, and $\hat{e}_o$ are
always mutually orthogonal, with $\hat{h}_o$ perpendicular to the
plane containing $\vec{v}_o$ and $\hat{r}$.

The paths of the extraordinary rays are less straightforward to find.
It is illustrative to pass from the Hamiltonian formulation for ray's
motion to a Lagrangian formulation.  This can be done by performing a
Legendre transformation on $\mathcal{H}_e$:
\begin{equation}
  \mathcal{L}_e(\vec{x},\Dot{\Vec{x}}) = \vec{p} \cdot \Dot{\Vec{x}} -
  \mathcal{H}_e, 
\end{equation}
where $\vec{p}$ can be written as a function of $\Dot{\Vec{x}}$ by
inverting the relation
\begin{equation}
  \Dot{\Vec{x}} = \nabla_{\vec{p}} \mathcal{H}_e =  (1 + \zeta) \vec{p}_e -
  \zeta (\vec{p}_e \cdot \hat{r})  \hat{r}. \label{extordv}
\end{equation}
(It can then be seen from \eqref{extordh},\eqref{extorde} and
\eqref{extordv} that $\vec{v}_e$, $\hat{h}_e$, and $\hat{e}_e$ are
mutually orthogonal, as is the case for the ordinary rays.)  To
perform this Legendre transform, we define an inverse metric tensor
$g^{ij}$ as
\begin{equation}
  g^{ij} = (1 + \zeta) \delta^{ij} -
  \zeta \hat{r}^i \hat{r}^j,
\end{equation}
in terms of which we have
\begin{equation}
  \mathcal{H}_e = \frac{1}{2} \left( g^{ij} p_i p_j - 1 \right).
\end{equation}
(The extra factors of $(1 + \zeta)$ in \eqref{extraordhamdef} were
chosen to make this inverse metric more tractable.)  It can then be
seen that $\Dot{x}^i = g^{ij} p_j$; inverting this, we then have $p_i
= g_{ij} \Dot{x}^i$, where $g_{ij}$ is the metric tensor itself:
\begin{equation}
  g_{ij} = \frac{1}{1 + \zeta} \delta_{ij} + \frac{\zeta}{1 + \zeta} \hat{r}_i
  \hat{r}_j.
\end{equation}
Performing the Legendre transformation on $\mathcal{H}_e$, then, we
find an effective point-particle Lagrangian for the extraordinary
rays:
\begin{equation}
  \mathcal{L}_e = \frac{1}{2} \left( g_{ij} \Dot{x}^i \Dot{x}^j + 1 \right),
\end{equation}
or, in terms of the path of the particle in spherical coordinates
$\{r(\tau), \theta(\tau), \phi(\tau)\}$,
\begin{equation}
  \mathcal{L}_e = \frac{1}{2} \left[ \dot{r}^2 + \frac{1}{1 + \zeta} r^2
    \left( \dot{\theta}^2 + \sin^2 \theta \dot{\phi}^2 \right) + 1 \right].
\end{equation}

This Lagrangian is independent of $\phi$ and $t$, implying that
there are two constants of the motion:  the angular momentum in the
$z$-direction, given by
\begin{equation}
  \ell =  \frac{1}{1 + \zeta} r^2 \sin^2 \theta \dot{\phi},
  \label{elldef}
\end{equation}
and the ``energy,''
\begin{equation}
  E = \frac{1}{2} \left[ \dot{r}^2 +  \frac{1}{1 + \zeta} r^2
    \left( \dot{\theta}^2 + \sin^2 \theta \dot{\phi}^2 \right) - 1 \right] = 0.
\end{equation}
Restricting our attention to the equatorial plane ($\theta =
\frac{\pi}{2}$), we can write this as
\begin{equation}  
  \frac{1}{2} \left[ \dot{r}^2 +
    (1 + \zeta) \frac{\ell^2}{r^2} - 1 \right] = 0
  \label{energycond}
\end{equation}
The motion of an extraordinary ray in the presence of a
Lorentz-violating monopole is therefore equivalent to the motion of a
particle with unit mass and total energy $E = 0$ in an effective
one-dimensional potential
\begin{equation}
  V_\text{eff}(r) = \frac{\ell^2 (1 + \zeta) }{2 r^2} - \frac{1}{2}.
\end{equation}

We now wish to calculate the trajectory of a ray originating far from
the monopole and passing nearby it.  Combining \eqref{elldef} and
\eqref{energycond}, we can write down a differential equation relating
$r$ and $\phi$ along the trajectory:
\begin{equation}
  \frac{d\phi}{dr} = \frac{\dot{\phi}}{\dot{r}}
  = \pm \frac{\ell (1 + \zeta) r^{-2}}{\sqrt{ 1 - \ell^2(1 + \zeta)
      r^{-2}}}.  
  \label{dphidr}
\end{equation}
The total angle of deflection $\Delta \phi$ of the light ray can then
in principle be found by integrating \eqref{dphidr} from $r = \infty$
to $r_\text{min}$ (defined as the value of $r$ for which $\dot{r} = 0$
in \eqref{energycond} above) and doubling the result.  However, a
closed-form analytic expression for $\Delta \phi$ cannot be found, for
the simple reason that $\zeta$ is a function of $r$, due to its
dependence on the field profile $g$; and as noted above, we do not
have a closed-form expression for $g(r)$.\footnote{Even if we wanted to
  evaluate this integral numerically using a numerical solution for
  the field profile, we would still expect the geometric-optics
  approximation to break down near the monopole core; in this region,
  the length scale of the field variation would presumably be shorter
  than the wavelength of the light waves being deflected.}  We can,
however, plug \eqref{gseries} into \eqref{bdef} to obtain a power
series for $\zeta$ in powers of $r^{-1}$.  Moreover, since $\xi b^2$
is expected to be many orders of magnitude less than one, we can
safely discard any terms of $\mathcal{O}(\xi^2 b^4)$ or higher.  All
told, then, we have the approximation
\begin{equation}
  \label{bseries}
  \zeta = \xi b^2 - \frac{\xi b^2}{2 \lambda
    b^2 r^2} + \mathcal{O}(\xi^2 b^4, r^{-4}).
\end{equation}
Defining $\mu = \xi b^2$ and $\nu = (2 \lambda b^2 \ell^2)^{-1}$, and
substituting $u = \ell/r$, we can write our integral for $\Delta \phi$
as
\begin{equation}
  \Delta \phi \approx \int_0^{u_\text{max}} \frac{1 + \mu - \mu \nu u^2}{\sqrt{1 - (1 +
      \mu) u^2 + \mu \nu u^4 }} du
\end{equation}
where $u_\text{max} = \ell/r_\text{min}$.  This quantity can then be
evaluated to leading order in $\mu$ and $\nu$, either by
differentiation with respect to $\mu$ and $\nu$ or in terms of
complete elliptic integrals.  To linear order in $\mu$ and $\nu$, the
result is
\begin{equation}
  \Delta \phi \approx \frac{\pi}{2} \left( 1 + \frac{\mu}{2} \left( 1
      - \nu \right) \right).
\end{equation}
The deflection angle $\alpha$ between the propagation direction of the
incoming ray and the outgoing ray will then be
\begin{equation}
  \alpha = 2 \Delta \phi - \pi \approx \frac{\pi}{2} \xi b^2 \left( 1
    - \frac{1}{2 \lambda b^2 \ell^2} \right),
  \label{alphadef}
\end{equation} 
where a positive value of $\alpha$ corresponds to a ray being
attracted towards the monopole, and a negative value corresponds to a
ray being repelled from the monopole.  Note that the sign of $\alpha$
is the same as the sign of $\xi$; the quantity in parentheses must be
assumed to be positive, since the quantity $\nu = (2 \lambda b^2
\ell^2)^{-1}$ has been assumed to be small.

It is instructive to ask what the meaning of the parameter $\nu$ is in
the above calculation.  To interpret it, we must assign some meaning
to the constant of the motion $\ell$; the most illuminating way to do
this is to relate it to the impact parameter $\beta$ of the ray.  The
asymptotic velocity of the ray (at $r \to \infty$) can be seen from
\eqref{energycond} to be $\dot{r} = \pm 1$; the plus or minus
corresponds to infalling or outgoing rays.  It can also be shown
geometrically that a particle at location $(x, y_0, 0)$ with velocity
$(-v, 0, 0)$ will have $\dot{\phi} = yv /r^2$.  Thus, the quantity
$\ell$ defined in \eqref{elldef} will be for this particle
\begin{equation}
  \ell = \frac{1}{1 + \zeta} y_0 v,
\end{equation}
In the limit $x \to \infty$ with $y_0$ fixed, we can identify $y_0$
with the impact parameter $\beta$ for this trajectory;  thus,
\begin{equation}
  \ell = \frac{1 - \frac{1}{6} \xi b^2}{1 + \frac{5}{6} \xi b^2} \beta,
\end{equation}
where the prefactor comes from taking the limit of $\zeta$ as $r \to
\infty$.

It is important to note that $\ell$ is not exactly equal to the impact
parameter in these units; however, since our result for the angular
deflection \eqref{alphadef} is only accurate to $\mathcal{O}(\xi^2
b^4)$, we can effectively replace $\ell$ with the impact parameter in
this equation:
\begin{equation}
  \alpha = 2 \Delta \phi - \pi \approx \frac{\pi}{2} \xi b^2 \left( 1
    - \frac{1}{2 \lambda b^2 \beta^2} \right),
  \label{alphabeta}
\end{equation} 
Note, meanwhile, that the characteristic length scale of the monopole
core (as found in \cite{Seifert2010}) is $r_M = (\sqrt{\lambda}
b)^{-1}$.  Thus, we have $\nu = \frac{1}{2}(r_M/\beta)^2$; in other
words, $\nu$ is best thought of as telling us about the ratio of the
physical size of the monopole to the impact parameter of the ray.

\section{Discussion}
\label{discsec}

We have shown that a direct coupling between the Maxwell field and an
antisymmetric tensor field will, in general, cause birefringent
effects.  Moreover, these effects are dependent on the local magnitude
of the field; this implies that a light ray will be deflected
depending on its polarization.  Light rays whose electric polarization
vector lies in the scattering plane (``ordinary rays'') will travel
past the monopole undeflected; those whose electric polarization is
perpendicular to the scattering plane, meanwhile (``extraordinary
rays'') will be deflected due to their varying speed of light.  The
leading-order angular deflection of these extraordinary rays, in terms
of the model's parameters $\xi$, $\lambda$, and $b$ and the ray's
impact parameter $\beta$ is given in \eqref{alphabeta}.

These effects are similar to the gravitational lensing effects derived
in previous work on the Lorentz-violating monopole solutions
\cite{Seifert2010, Seifert2010a, Li2012}, although there are some
important differences as well.  In the work of Li \emph{et al.}\
\cite{Li2012}, the gravitational deflection of a null ray due to a
Lorentz-violating monopole solution is shown to be
\begin{equation}
  \alpha_G = \frac{3}{2} \pi \epsilon -
  \frac{\sqrt{\epsilon}}{\sqrt{\lambda} b r_m} + \frac{\epsilon}{20
    \lambda b^2 r_m^2} + \mathcal{O}(r_m^3),
  \label{alphagrav}
\end{equation}  
where $\epsilon = 16 \pi G b^2$ and $r_m$ is the radial coordinate of
closest approach for the null ray in question.  The most notable
similarity between the gravitational deflection $\alpha_G$ and the
direct-coupling deflection $\alpha$ is that in both cases, the
deflection angle does not vanish in the limit of large impact
parameter.  (This is to be contrasted with the case of light
deflection by a conventional Schwarzchild metric, for which the
deflection angle goes to zero as the impact parameter gets large.)  In
our case, we have
\begin{equation}
  \alpha_\infty \equiv \lim_{\beta \to \infty} \alpha = \frac{\pi}{2}
  \xi b^2;
\end{equation}
the corresponding quantity in the gravitational case is
$(\alpha_G)_\infty = \frac{3}{2} \pi \epsilon$.  In this asymptotic
limit, light rays behave as though there was a conical deficit angle
due to the monopole.

However, this observation also illuminates an important difference
between the direct-coupling case and the gravitational case.  As noted
previously, the coupling parameter $\xi$ can be either positive or
negative; from \eqref{alphadef}, this implies that extraordinary rays
are attracted towards the monopole when $\xi > 0$, but are repelled by
it when $\xi < 0$.  In the limit $\beta \to \infty$, the light rays
behave as if in a space with a conical deficit angle when $\xi$ is
positive, and a ``conical surplus angle'' when $\xi$ is negative.  In
contrast, light rays are always deflected towards a monopole solution
by its gravitational influence, since $\epsilon$ (and therefore
$\alpha_G$, in the regime where \eqref{alphagrav} is valid) is always
positive.

\begin{figure}[tb]
  \begin{tabular*}{\columnwidth}{c @{\extracolsep{\fill}} c}
    \includegraphics[width=0.48\columnwidth]{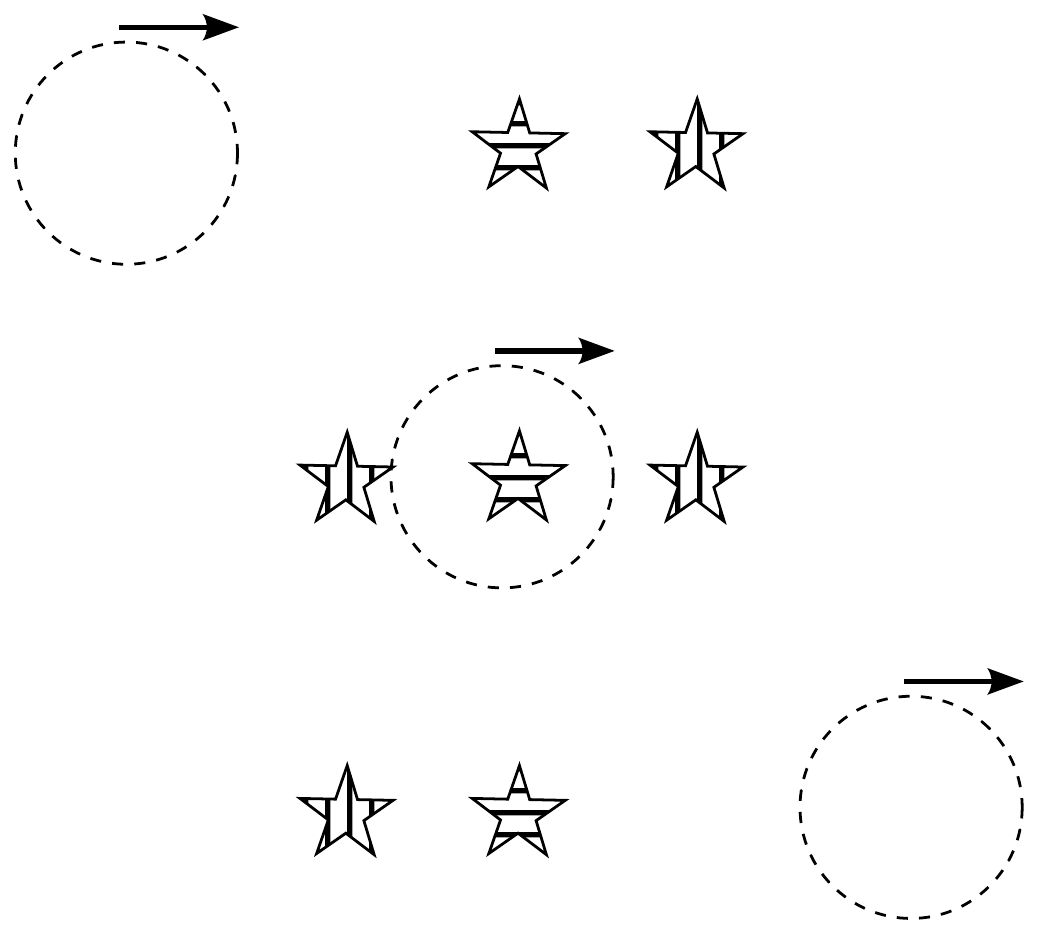}
    & \includegraphics[width=0.48\columnwidth]{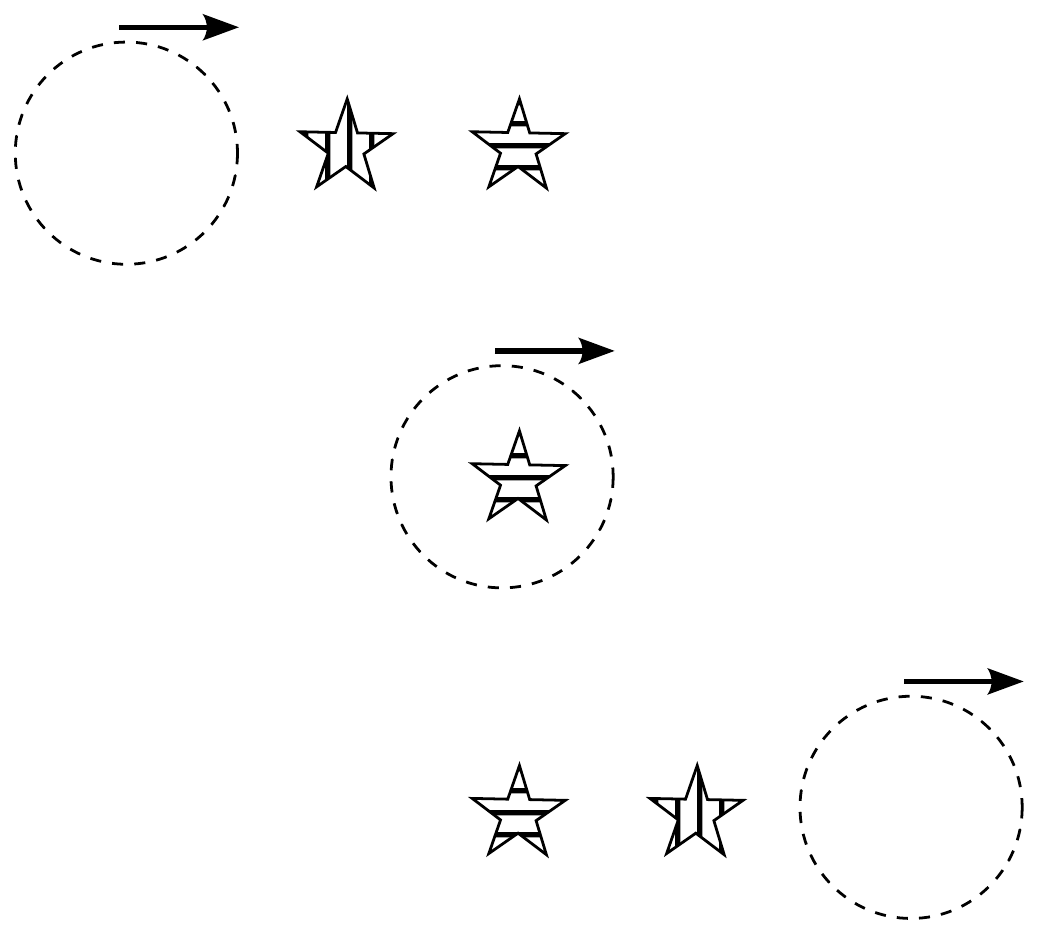} 
    \\
    (a) & (b) \\
  \end{tabular*}
  \caption{\label{monopolefig} Multiple images of a distant star
    created by a passing monopole with (a) $\xi >0$; (b) $\xi < 0$.
    The monopole's location is indicated by the dashed circle; the
    (electric) polarization of the light from each image is indicated
    by the direction of its stripes.}
\end{figure}

We can also ask what we would see if a monopole lay between us and a
distant star, in the case where $\xi \neq 0$ but neglecting the
gravitational effects on the light rays.  For simplicity, consider the
case in which the ray trajectory stays far from the monopole, and so
the approximation $\nu \approx 0$ is valid.  It will always be the
case that the ordinary rays sent out by the star will reach us along a
straight-line path; the image observed this way would be highly
polarized, but otherwise undistorted.  However, the extraordinary rays
would form zero, one, or two images, depending on the sign of $\xi$
and the angular separation between the monopole and the distant star
on the sky.  A schematic illustration of the images of a distant star
created by a passing monopole is shown in Figure \ref{monopolefig}.

Of particular note is that there is \emph{always} a double image of
the star due to the birefringent effects of the monopole; the angular
separation of the ordinary image and extraordinary image (in the limit
$\nu \to 0$) will be 
\begin{equation} 
  \theta_E = \frac{\pi}{2} \xi b^2 \frac{D_{MS}}{D_S},
\end{equation}
where $D_{MS}$ is the distance to the source from the monopole and
$D_S$ is the distance to the star.\footnote{The angle $\theta_E$ would
  also be the radius of the ``Einstein ring'' in the case of perfect
  alignment between Earth, the monopole, and the distant star.}  Given
the non-observation of such double images on the sky, we can
immediately say that if $\xi b^2$ is non-zero, it must be sufficiently
small that these double images are not observed.  If we take the
maximum angular resolution of the Very Long Baseline Array (about $1.2
\times 10^{-4}$ arcseconds \cite{RomneyJD}) as the best achievable
resolution with current technology, we must thereby conclude that
$|\xi b^2| < 10^{-10}$ or so.  This is, of course, a much less
stringent bound than those placed on $\xi b^2$ via direct polarimetry
measurements (see Section \ref{paramboundsec}); we therefore conclude
that these multiple images will, in general, be so closely spaced as
to be unresolvable. However, the effects of this coupling would still
in principle be observable via intensity variations of the starlight,
either due to the appearance and disappearance of the aforementioned
multiple images, or via more conventional microlensing effects.  We
are currently investigating these possibilities.

Finally, we note that the coupling \eqref{kdef} is not the only
possible coupling between an antisymmetric tensor field $B_{ab}$ and
the Maxwell field $A_a$.  In particular, one could also couple
$B_{ab}$ to the Maxwell field by postulating an ``effective metric''
$\tilde{\eta}^{ab}$, differing from the ``canonical'' metric
$\eta^{ab}$:
\begin{equation}
  \tilde{\eta}^{ab} = \eta^{ab} + \chi B^a {}_c B^{bc}
  \label{effmetricdef}
\end{equation}
The Lagrangian for a Maxwell field that operates under this metric instead
of $\eta^{ab}$ would then be
\begin{equation}
  \mathcal{L} = - \frac{1}{4} \tilde{\eta}^{ac} \tilde{\eta}^{bd}
  F_{ab} F_{cd} = - \frac{1}{4} \left( F_{ab} F^{cd} + (\tilde{k}_F)^{abcd} F_{ab}
  F_{cd} \right),
\end{equation}
where
\begin{multline}
  \label{tildekdef}
  (\tilde{k}_F)^{abcd} = \frac{\chi}{2} \left[ B^{[a} {}_e \eta^{b][c}
    B^{d]e} + \frac{2}{3} \eta^{a[c} \eta^{d]b} B_{ef} B^{ef} \right]
  \\ +
  \mathcal{O}(\chi^2 b^4).
\end{multline}
The effects of this coupling could presumably be analysed using the
same methods in this work.  Importantly, however, this coupling should
not lead to birefringent effects (for the same reasons that the
``generalized bumblebee models'' analysed in \cite{Seifert2010b} did
not lead to birefringence.)  The parameter $\chi$ in such a model
would therefore evade the stringent constraints that were imposed by
polarimetry on the parameter $\chi$ in our original coupling
\eqref{kdef}.

\acknowledgments

We thank Matthew Mewes and Brett Altschul for helpful discussions
during the preparation of this work.

\appendix*

\section{Dependence of $\sigma$ on model parameters}

Following the language of \cite{Kostelecky2002}, the parameter
$\sigma$ for a source with right ascension $\alpha$ and declination
$\delta$ is given by
\begin{equation}
  \sigma^2 = (\vec{\varsigma}_a \cdot \vec{k})^2 + (\vec{\varsigma}_c
  \cdot \vec{k})^2, 
\end{equation}
where we have defined the three ten-dimensional vectors in this
equation as

\begin{equation}
  \varsigma_s^a = \begin{bmatrix}
    \cos^2 \delta + \cos^2 \alpha - \sin^2 \alpha \sin^2 \delta \\
    \sin^2 \delta \cos^2 \alpha - \cos^2 \delta - \sin^2 \alpha \\
    - 2 \sin \delta \sin \alpha \cos \alpha \\
    - \sin \delta \sin \alpha \cos \alpha \\
    \sin \delta (\sin^2 \alpha - \cos^2 \alpha) \\
    - \cos \delta \sin \alpha \\
    \cos \delta \cos \alpha \\
    - \sin \delta \cos \delta \cos \alpha \\
    - \cos^2 \delta \sin \alpha \cos \alpha \\
    - \sin \delta \cos \delta \sin \alpha
  \end{bmatrix},
\end{equation}
\begin{equation}
  \varsigma^a_c = \begin{bmatrix}
    - 2 \sin \delta \sin \alpha \cos \alpha \\
    - 2 \sin \delta \sin \alpha \cos \alpha \\
    \frac{1}{2} (1 + \sin^2 \delta) (\sin^2 \alpha - \cos^2 \alpha) \\
    \frac{1}{2} (\sin \delta + \sin^2 \alpha - \sin^2 \delta \cos^2
    \alpha ) \\
    (1 + \sin^2 \delta) \sin \alpha \cos \alpha \\
    - \sin \delta \cos \delta \cos \alpha \\
    - \sin \delta \cos \delta \sin \alpha \\
    \cos \delta \sin \alpha \\
    \sin \delta (\sin^2 \alpha - \cos^2 \alpha) \\
    - \cos \delta \cos \alpha 
  \end{bmatrix},
\end{equation}
and
\begin{equation}
  k^a = \begin{bmatrix} - \bar{Q}^y \bar{R}^y + \frac{1}{3}
    \bar{Q} \cdot \bar{R} \\
    \bar{Q}^x \bar{R}^x - \frac{1}{3}
    \bar{Q} \cdot \bar{R} \\
    (\Bar{Q}^y)^2 - (\Bar{R}^y)^2 - \frac{1}{3} \left( \bar{Q}^2 -
      \bar{R}^2 \right) \\
    (\Bar{Q}^z)^2 - (\Bar{R}^z)^2 - \frac{1}{3} \left( \bar{Q}^2 -
      \bar{R}^2 \right) \\
    \bar{Q}^x \bar{Q}^y - \bar{R}^x \bar{R}^y \\
    \bar{Q}^x \bar{Q}^z - \bar{R}^x \bar{R}^z \\
    \bar{Q}^y \bar{Q}^z - \bar{R}^y \bar{R}^z \\
    \bar{Q}^x \bar{R}^z + \bar{Q}^z \bar{R}^x \\
    - \bar{Q}^x \bar{R}^y - \bar{Q}^y \bar{R}^x \\
    \bar{Q}^y \bar{R}^z + \bar{Q}^z \bar{R}^y \end{bmatrix}.
\end{equation}
Note that this last vector contains the ten independent parameters of
the matrices $\tilde{\kappa}_{e+}$ and $\tilde{\kappa}_{o-}$.
 
\bibliographystyle{./custombib}
\bibliography{./Monopole_lensing}{}

\begin{thebibliography}{10}
\newcommand{\enquote}[1]{``#1''}
\providecommand{\url}[1]{\texttt{#1}}
\providecommand{\urlprefix}{URL }
\providecommand{\eprint}[2][]{\url{#2}}

\bibitem{Colladay1998}
Colladay, D. and Kosteleck\'{y}, V.~A., Physical Review D, \textbf{58}, 116002
  (1998).

\bibitem{Kostelecky2008}
Kostelecký, V.~A. and Russell, N., \enquote{{Data Tables for Lorentz and CPT
  Violation}.} (2008), \eprint{ArXiv:0801.0287v6}.

\bibitem{Kostelecky2004}
Kosteleck\'{y}, V.~A., Physical Review D, \textbf{69}, 105009 (2004).

\bibitem{Bluhm2005}
Bluhm, R. and Kosteleck\'{y}, V.~A., Physical Review D, \textbf{71}, 065008
  (2005).

\bibitem{Bluhm2008}
Bluhm, R., Fung, S.-H., and Kosteleck\'{y}, V.~A., Physical Review D,
  \textbf{77}, 065020 (2008).

\bibitem{Seifert2009}
Seifert, M.~D., Physical Review D, \textbf{79}, 124012 (2009).

\bibitem{Seifert2010b}
Seifert, M.~D., Physical Review D, \textbf{81}, 065010 (2010).

\bibitem{Kostelecky2009}
Kosteleck\'{y}, V.~A. and Potting, R., Physical Review D, \textbf{79}, 065018
  (2009).

\bibitem{Altschul2010}
Altschul, B., Bailey, Q.~G., and Kosteleck\'{y}, V.~A., Physical Review D,
  \textbf{81}, 065028 (2010).

\bibitem{Seifert2010}
Seifert, M.~D., Physical Review Letters, \textbf{105}, 201601 (2010).

\bibitem{Seifert2010a}
Seifert, M.~D., Physical Review D, \textbf{82}, 125015 (2010).

\bibitem{Kibble1976}
Kibble, T., Journal of Physics A: Mathematical and General, \textbf{9}, 1387
  (1976).

\bibitem{Kostelecky2002}
Kosteleck\'{y}, V.~A. and Mewes, M., Physical Review D, \textbf{66}, 056005
  (2002).

\bibitem{Kostelecky2001}
Kosteleck\'{y}, V.~A. and Mewes, M., Physical Review Letters, \textbf{87},
  251304 (2001).

\bibitem{Kostelecky2013}
Kosteleck\'{y}, V.~A. and Mewes, M., Physical Review Letters, \textbf{110},
  201601 (2013).

\bibitem{Li2012}
Li, X.-Z., Xi, P., and Zhang, Q., Physical Review D, \textbf{85}, 085030
  (2012).

\bibitem{Sluijter2008}
Sluijter, M., de~Boer, D. K.~G., and Braat, J. J.~M., Journal of the Optical
  Society of America. A, Optics, image science, and vision, \textbf{25}, 1260
  (2008).

\bibitem{RomneyJD}
National Radio Astronomy Observatory, \emph{{VLBA Observational Status Summary
  2013B}} (2013), available at \texttt{science.nrao.edu/facilities/vlba/docs/}.

\end{thebibliography}

\end{document}